\documentclass[twocolumn,epjc3]{svjour3}
\usepackage[english]{babel}
\usepackage{amssymb}
 \usepackage{amsmath}
\usepackage{latexsym}
\usepackage{epsfig}
\usepackage{color}
\usepackage{float}
\usepackage{graphicx}
\usepackage{epstopdf}
\usepackage[square,comma,numbers,sort&compress]{natbib}

\usepackage[hidelinks=true]{hyperref}
\hypersetup{
  colorlinks   = true, 
  urlcolor     = red, 
  linkcolor    = red, 
  citecolor   = blue 
}
\journalname{Eur. Phys. J. C}
\begin{document}

\title{\Large Observational constraints on interacting Tsallis holographic dark energy model}
\author{Ehsan Sadri\thanksref{e1,addr1}}
\thankstext{e1}{E-mail: ehsan@sadri.id.ir}
\institute{Azad University Central Tehran Branch, Tehran 34353-17117, Iran\label{addr1}}

\maketitle
\abstract{
In this paper, we investigate a recent proposed model - so called the Tsallis holographic dark energy (THDE) model with consideration of the Hubble and the event future horizon as IR cutoffs. In this case, we consider the non-gravitational and phenomenological interaction between dark sectors. We fit the free parameters of the model using Pantheon Supernovae Type Ia data, Baryon Acoustic Oscillations, Cosmic Microwave Background, Gamma-Ray burst and the the local value of the Hubble constant. We examine the THDE model to check its compatibility with observational data using objective Information Criterion (IC). We find that the THDE models cannot be supported by observational data once the $\Lambda$CDM is considered as the referring model. Therefore we re-examine the analysis with the standard holographic dark energy model (HDE) as another reference.  Changing the $\Lambda$CDM to main standard dark energy model (HDE), we observe the compatibility of the THDE models. Using the Alcock-Paczynski (AP) test we check the deviation of the model compared to $\Lambda$CDM and HDE. Surveying the evolution of squared of sound speed $v^2_s$ as an another test we check the stability of the interacting and non-interacting THDE models and we find that while the THDE model with the Hubble horizon as IR cutoff is unstable against the background perturbation, the future event horizon as IR cutoff show stability at the late time. In addition, using the modified version of the CAMB package, we observe the suppressing the CMB spectrum at small K-modes and large scale.}


\section{Introduction}
\indent Dark energy - raised in 1998 - with strong negative pressure is the main component of the acceleration of the universe\citep{riess,garnavich,perlmutter,riess2,tegmark,spergel,Calabrese,ywang,mzhao}. The nature of dark energy has become one of the main unknown issues in modern cosmology and many models have been proposed to understand this new concept\citep{bamba,lili,wang,Peebles,padman,Copeland,Mli}. The cosmological constant as the simplest dark energy fluid is a good candidate to study the universe acceleration and behavior of the dark energy. However, the cosmological constant suffers from some problem\citep{Weinbergbook,Sahni,Garriga,Frieman,Nojiri,caldwell}. In order to alleviate these problems a holographic model of dark energy has been suggested and has drawn many attentions in recent two decades\citep{wanggong,liwang,itom,weih,nojiri1,nojiri2,nojiri3,nojiri4}. As the name of holographic dark energy model (HDE) suggests, this model is originated from holographic principle and its energy density can be expressed by $\rho_D=3c^2M^2_P/L^2$ where $c^2$ is a numerical constant, $M_P$ is the reduced Planck mass and $L$ denotes the size of the current universe such as the Hubble scale\citep{horava,thomas}. In addition, the HDE has some problems and cannot explain the timeline of a flat FRW universe\citep{shsu,mili}. One of the proposed solutions for the HDE problems is the consideration of different entropies. One of the considered entropies is Tsallis entropy which has been used in many works\citep{te1,te2,te3,te4,majh,abe,touch,biro}.\\
\indent In the recent work, using the concept of holography in the Tsallis entropy a new holographic dark energy models is appeared so-called the Tsallis holographic dark energy model(THDE)\citep{thde}. It is stated that by applying the Tsallis statics\citep{majh,abe,touch,biro} to the system horizon the Bekenstein entropy can be achieved and leads to stable models\citep{jah}. The authors with choosing the Hubble horizon as IR cutoff studied different aspects of the THDE model without consideration of interaction between dark sectors \citep{thde}. The model has been studied to be checked if it can satisfy the condition of FRW universe and it has been found that the non-interacting model is unstable. On the other hand, another work reformulated the Tsallis HDE with the future event horizon as IR cutoff and explained that the Tsallis HDE with consideration of the Hubble horizon as IR cutoff does not embrace the standard HDE and consequently standard thermodynamic which can be considered as a substantial drawback\citep{thde2}.\\
\indent In this work, in the direction of the two mentioned works, we would like to investigate the behavior of the Tsallis Holographic Dark Energy model (THDE) with consideration of a non-gravitational and phenomenological interaction. In addition we choose both cutoffs namely, the Hubble horizon and the future event horizon for figuring out the comparison between them. This will be done by the use of the latest observational data, namely the Pantheon Supernovae type Ia, Baryon Acoustic Oscillations (BAO), Cosmic Microwave Background (CMB), Gamma-Ray burst and the the local value of the Hubble constant. We check the compatibility of the models with observational data employing AIC and BIC model selection tools. In addition, using Alcock-Pecinski (AP) test we survey the deviation of the THDE from the $\Lambda$CDM as the reference model and we make a comparison with the well-known holographic dark energy (HDE) model. Using the best values of the models' free parameters we check the stability of the interacting THDE. We also study the behavior of the model in CMB angular power spectrum.\\
\indent The structure of this paper is as follows. In the next section (section 2) we introduce the background physics of the THDE with consideration of an interaction between dark sectors. In section 3, we introduce the data and method used in this work. In section 4, we discuss the results of data analysis. In section 5, we study the AP test for measuring the deviation of THDE compared to the HDE and $\Lambda$CDM models. In section 6, we study the evolution of stability of the model within the different redshift values. The section 7 is allocated to the behavior of the THDE in the CMB angular power spectrum. The last section is dedicated to some concluding remarks.
\section{Background of THDE}
The description of a homogeneous and isotropic Friedmann-Lema\^itre-Roberston-Walker (FRW) universe can be introduce by,
\begin{equation} \label{flrw}
ds^2=-dt^2+a^2\left(t\right)\left(\frac{dr^2}{1-kr^2}+r^2d\Omega^2\right)\,,
\end{equation}
in which $a$ is the scale factor and $k=0,1,-1$ denote a flat, closed and open universe, respectively . For a spatially flat FRW universe the first Friedmann equations can be written as
\begin{equation}\label{frw}
H^2=\frac{1}{3M_P^2}(\rho_D+\rho_m),
\end{equation}
where $\rho_D$ is the THDE energy density, $\rho_m$ is the energy density of pressureless matter which must contain all the material constituents of the Universe and $M_p$ denotes the reduced Planck mass. We may also write the dark energy and dark matter density with respect to the critical density $\rho_{cr}=3M^2_pH^2$ as
\begin{equation}\label{omgdcr}
\Omega_{D}=\frac{\rho_{D}}{3M_P^2 H^2},
\end{equation}
\begin{equation}\label{omgmcr}
\Omega_{m}=\frac{\rho_{m}}{3M_P^2 H^2},
\end{equation}
The energy density of the Tsallis holographic dark energy (THDE) is given by the following relation
\begin{equation}\label{THDE1}
\rho_{D}=BL^{2\delta-4}\,,
\end{equation}
where $B$ is an unknown parameter, $L$ can be considered as the size of the current Universe such as the Hubble scale and  the future event horizon $\delta$ is a free parameter reduces the THDE to the HDE model at $\delta=1$.
\subsection{THE HUBBLE HORIZON}
Using the Hubble horizon as the IR cutoff of the system $L=H^{-1}$, the Eq.(\ref{THDE1}) takes the form
\begin{equation}\label{THDE2}
\rho_{D}=BH^{4-2\delta}\,,
\end{equation} 
Taking time derivative of the equation above and using Eq.(\ref{omgdcr}), we reach
\begin{equation}\label{THDE3}
\dot{\Omega}_D=\left(-2\delta+2\right)\Omega_D\frac{\dot{H}}{H},
\end{equation} 
Considering an interaction term between dark sectors, one can write the conservation equations for THDE as
\begin{equation}\label{cons1}
\dot{\rho}_m+3H\rho_m=Q,
\end{equation}
\begin{equation}\label{cons2}
\dot{\rho}_D+3H(\rho_D+P_D)=-Q,
\end{equation}
where $Q$ indicates the interaction term explaining energy flow between the components. Regarding the usual option for $Q$-term as $3H(b_1\rho_D+b_2\rho_m)$ which $b_{1,2}$ are the coupling constant, a single coupling constant can be used properly (e.g. $Q_1=3Hb\rho_D$, $Q_2=3Hb\rho_m$ and $Q_3=3Hb(\rho_D+\rho_m)$. In our recent work, we compared different phenomenological linear and non-linear interaction cases in the framework of the holographic Ricci dark energy model \cite{skz} and we found that the linear interaction $Q=3Hb\rho_D$ and $Q=3Hb\rho_m$ are the best cases among the others. Accordingly, in this work we take $Q=3Hb\rho_m$ in our calculations.\\
Taking time derivative of Eq.(\ref{frw}) and using Eqs.(\ref{frw}), (\ref{cons1}), (\ref{cons2}) we find a definition for dark energy pressure
\begin{equation}\label{p}
P_D=-\frac{2}{3}\frac{\dot{H}}{H^2}\left(\rho_D+\rho_m\right)-\rho_D-\rho_m,
\end{equation}
combining the Eqs.(\ref{p}) and (\ref{cons2}) yields
\begin{equation}\label{omgh}
\dot{\Omega}_D+3H\left(-\frac{2}{3}\frac{\dot{H}}{H^2}\right)-3H\Omega_m=0,
\end{equation}
Inserting the Eq.(\ref{THDE3}) into Eq.(\ref{omgh}) leads to
\begin{equation}\label{hh}
\frac{\dot{H}}{H^2}=3\frac{1-\Omega_D+3b\Omega_m}{2\Omega_D(2-\delta)-2},
\end{equation}
and combining the  Eqs.(\ref{hh}) and (\ref{THDE3}) we have
\begin{equation}\label{omgd}
\dot{\Omega}_D=6(1-\delta)\Omega_D\left(\frac{1-\Omega_D+3b\Omega_m}{2\Omega_D(2-\delta)-2}\right),
\end{equation}
in which $\dot{\Omega}_D=\Omega'_DH$ and $\dot{H}=H'H$ where the prime denotes the derivative respect to $x=lna$ and $a=(1+z)^{-1}$. Then the evolution of the density of dark energy and the Hubble parameter for THDE in terms of redshift can be written as
\begin{equation}\label{omg}
\frac{d\Omega_D}{dz}=-\left(\frac{1}{1+z}\right)\left(6(1-\delta)\Omega_D\left(\frac{1-\Omega_D+3b\Omega_m}{\Omega_D(4-2\delta)-2}\right)\right),
\end{equation}
\begin{equation}\label{hhd}
\frac{dH}{dz}=\left(\frac{H}{1+z}\right)\left(3\frac{1-\Omega_D+3b\Omega_m}{\Omega_D(4-2\delta)-2}\right).
\end{equation}
in which $\Omega_m=1-\Omega_D$, $b$ is the coupling constant and $\delta$ is the free parameter of THDE model.
\subsection{THE FUTURE EVENT HORIZON}
Now, using the future event horizon as the IR cutoff of the system $L=R_h$ 
\begin{equation}\label{Rh}
R_h=a\int_a^\infty \frac{da}{Ha^2},
\end{equation}
the Eq.(\ref{THDE1}) will be
\begin{equation}\label{THDErh}
\rho_{D}=BR_h^{4-2\delta}\,,
\end{equation} 
Using the Eqs.(\ref{omgmcr}), (\ref{cons1}), (\ref{Rh}), (\ref{THDErh}) and $Q=3bH\rho_m$ one can reach the following equations expressing the evolution of the density of dark energy and the Hubble parameter 
\begin{equation}\label{omgRh}
\begin{split}
&\frac{d\Omega_D}{dz}=-\left(\frac{1}{1+z}\right)\Omega_D(\Omega_D-1)\Big(2 (\delta -2)\\
&\times\sqrt{-a^{-3 (b-1) } (\Omega_D-1)} \left(\frac{ a^{3 (b-1) } (\Omega_D-1) \Omega_D}{(\Omega_D-1)/H_0^2}\right)^{\frac{1}{4-2 \delta }}\\
&+H_0 (3 b-2 \delta +1) \sqrt{1-\Omega_D}\Big)\\
&\times\left(H_0 \sqrt{1-\Omega_D}\right)^{-1}
\end{split}
\end{equation} 
\begin{equation}\label{HRh}
H=\frac{H_0\sqrt{\Omega_{m0}a^{3(b-1)}}}{\sqrt{1-\Omega_D}}.
\end{equation} 
in which $a=(1+z)^{-1}$. It is possible to write the Eq.(\ref{HRh}) as the following form
 \begin{equation}\label{HRhp}
\begin{split}
&\frac{dH}{dz}=-\left(\frac{H}{1+z}\right)\Big(\frac{H_0}{2}\sqrt{1-\Omega_{D_0}}(1+z)^{3(b-1)}\\
&\left(3(1-b)(\Omega_D-1)+\Omega'_D\right)\times\left((1+z)^{3(b-1)}(1-\Omega_D)\right)^{-\frac{3}{2}}
\end{split}
\end{equation} 
where $a=(1+z)^{-1}$ and $H_0$ is the present value of the Hubble parameter (the Hubble constant), $\Omega_{D_0}$ is the present value of the density of dark energy.
\section{Observational Data}
To analyze the models and to obtain the best fit values for the model parameters, in this paper we combine the latest observational data including BAO, CMB, SNIa, $H_0$ and GRB. For this purpose, we employed the public codes EMCEE \cite{emcee} for implementing the MCMC method and GetDist Python package\footnote{https://getdist.readthedocs.io} for analyzing and plotting the contours.\\
\subsection{Supernovae Type Ia}\label{snpa}
The compilation of Pantheon sample including 1048 data points embrace the redshift range $0.01<z<2.3$ \citep{pant}. This sample contains 276 SNIa case from PanSTARRS1 Medium Deep Survey, SDSS, Low$-z$ and HST samples. We use the systematic covariance $C_{sys}$ for a vector of binned distances
\begin{equation}\label{SNsys}
C_{ij,sys}=\sum_{n=1}^{i}\left(\frac{\partial \mu_i}{\partial S_n}\right)\left(\frac{\partial \mu_j}{\partial S_n}\right)\left(\sigma_{S_k}\right)
\end{equation}
in which the summation is over the $n$ systematic with $S_n$ and its magnitude of its error $\sigma_{S_n}$. The $\chi^2$ relation for Pantheon SNIa data is
\begin{equation}\label{SNchi2}
\chi^2_{Pantheon}=\triangle\mu^T\cdot C_{Pantheon}^{-1}\cdot\triangle\mu
\end{equation}
in which  $\triangle\mu=\mu_{data}-M-\mu_{obs}$ and $M$ is a nuisance parameter. It should be note that the $C_{Pantheon}$ is the summation of the systematic covariance and statistical matrix $D_{stat}$ having a diagonal component. The complete version of full and binned Pantheon supernova data are provided in the online source\footnote{https://archive.stsci.edu/prepds/ps1cosmo/index.html}\\
\subsection{Baryon Acoustic Oscillations}
The combination of the extended Baryon Oscillation Spectroscopic Survey (eBOSS) quasar clustering at $z=1.52$ \citep{eboss}, isotropic BAO measurements of 6dF survey at an effective redshift ($z=0.106$) \citep{6df} and the BOSS DR12 \cite{dr12} including six data points of Baryon Oscillations as the latest observational data for BAO makes the total data used for BAO in this section. The $\chi^2_{BAO}$ of BOSS DR12 can be explained as
\begin{equation}\label{BOSSDR12}
\chi^2_{BOSS~DR12}=X^tC_{BAO}^{-1}X,
\end{equation}
where $X$ for six data points is
\begin{equation}\label{XBAO}
X=\left(\begin{array}{c} \frac{D_M\left(0.38\right)r_{s,fid}}{r_s\left(z_d\right)}-1512.39\\
\frac{H\left(0.38\right)r_s\left(z_d\right)}{r_s\left(z_d\right)}-81.208\\
\frac{D_M\left(0.51\right)r_{s,fid}}{r_s\left(z_d\right)}-1975.22\\
\frac{H\left(0.51\right)r_s\left(z_d\right)}{r_s\left(z_d\right)}-90.9\\
\frac{D_M\left(0.61\right)r_{s,fid}}{r_s\left(z_d\right)}-2306.68\\
\frac{H\left(0.51\right)r_s\left(z_d\right)}{r_s\left(z_d\right)}-98.964\end{array}\right),
\end{equation}
and $r_{s,fid}=$147.78 Mpc is the sound horizon of fiducial model, $D_M\left(z\right)=\left(1+z\right)D_A\left(z\right)$ is the comoving angular diameter distance. The sound horizon at the decoupling time $r_s\left(z_d\right)$ is defined as

\begin{equation}\label{BAO}
r_s\left(z_d\right)=\int_{z_d}^{\infty} \frac{c_s\left(z\right)}{H\left(z\right)}dz,
\end{equation}
in which $c_s=1/\sqrt{3\left(1+R_b/\left(1+z\right)\right)}$ is the sound speed with
$R_b=31500\Omega_bh^2\left(2.726/2.7\right)^{-4}$. The covariance matrix $Cov_{BAO}$ \cite{dr12} is:

The $\chi^2$ for combined data is
\begin{equation}\label{BAO}
\chi^2_{BAO}=\chi^2_{BOSS~DR12}+\chi^2_{6dF}+\chi^2_{eBOSS}.
\end{equation}
\subsection{Cosmic Microwave Background}
Surveying the evolution of the expansion history of the universe leads us to check the Cosmic Microwave Background (CMB). For this, we use the data of Planck 2015 \cite{cmb1}. The $\chi^2_{CMB}$ function may be explained as
\begin{equation}\label{CMB}
\chi^2_{CMB}=q_i-q^{data}_i Cov^{-1}_{CMB}\left(q_i,q_j\right),
\end{equation}
where $q_1=R\left(z_*\right)$, $q_2=l_A\left(z_*\right)$ and $q_3=\omega_b$ and $Cov_{CMB}$
is the covariance matrix \cite{cmb1}. The data of Planck 2015 are
\begin{equation}\label{PLANCKDATA}
q^{data}_1=1.7382,~\\
q^{data}_2=301.63,~\\
q^{data}_3=0.02262.
\end{equation}
The acoustic scale $l_A$ is
\begin{equation}\label{lA}
l_A=\frac{3.14d_L\left(z_*\right)}{\left(1+z\right)r_s\left(z_*\right)},
\end{equation}
in which $r_s\left(z_*\right)$ is the comoving sound horizon at the drag epoch ($z_*$). The function of redshift at the drag epoch is \cite{cmb2}

\begin{equation}\label{z_*}
z_*=1048\left[1+0.00124\left(\Omega_bh^2\right)^{-0.738}\right]\left[1+g_1\left(\Omega_mh^2\right)^{g_2}\right],
\end{equation}
where
\begin{equation}\label{g1 g2}
g_1=\frac{0.0783\left(\Omega_bh^2\right)^{-0.238}}{1+39.5\left(\Omega_bh^2\right)^{-0.763}}, ~~~g_2=\frac{0.560}{1+21.1\left(\Omega_bh^2\right)^{1.81}}.
\end{equation}
The CMB shift parameter is \cite{cmb3}
\begin{equation}\label{R}
R=\sqrt{\Omega_{m_0}}\frac{H_0}{c}r_s\left(z_*\right).
\end{equation}
The reader should notice that the usage of CMB data does not provide the full Planck information but it is an optimum way of studying wide range of dark energy models.\\ 
\subsection{Gamma-Ray Burst} 
Constraining the free parameters using Gamma-Ray burst data can be obtained by fitting the distance modulus $\mu(z)$ similar to SNIa data (Sec.\ref{snpa}). In this work we use 109 data of Gamma-Ray Burst in the redshift range $0.3<z<8.1$\citep{grbref}. This data contains 50 low-z GRBs ($z<1.4$) and high-z GRBs ($z>1.4$). The 70 GRBs are obtained in \citep{amat1}, 25 GRBs are taken from\citep{amat2} and the rest 14 GRBs data points are provided from \citep{amat3}. The $\chi^2$ for GRB is given by
\begin{equation}\label{grbfunc}
\chi^2_{GRB}=\sum_{i=1}^{109} \frac{\left[\mu_{obs}(z_i)-\mu_{th}(z_i)\right]^2}{\sigma^2(z_i)},
\end{equation}
in which the theoretical distance modulus $\mu_{th}(z_i)$ can be defined as
\begin{equation}\label{grbfunc}
\mu_{th}(z_i)=5log_{10}D_l(z_i)+\mu_0.
\end{equation}
where $\mu_0=42.38-5log_{10}h$ and $h=H_0/100$ with unit of km/s/Mpc.
\subsection{Local Hubble Constant}
As the last data point, we use the $H_0$ which can be locally measured by ladder distance observation.  According to the reported result \citep{localh} we use $H_0=73.24\pm1.74$ $kms^{-1}/Mpc^{-1}$ in our analysis.\\

\begin{figure*}[!]
   \centering
\begin{tabular}{cc}
\hspace*{-0.1in}
\includegraphics[width=0.45\textwidth]{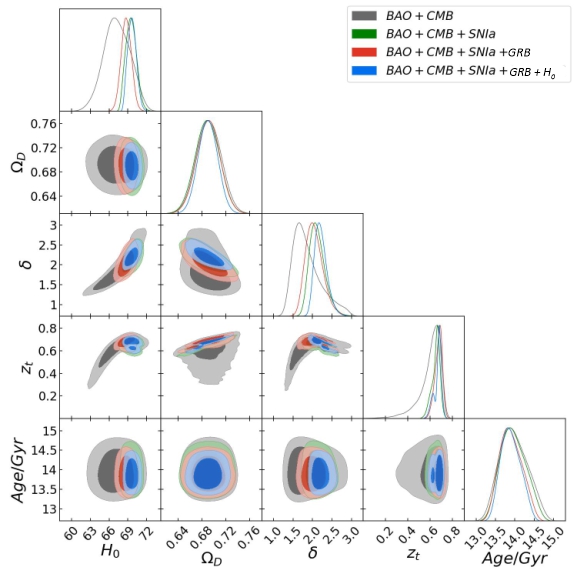}\hspace{6mm}
\includegraphics[width=0.45\textwidth]{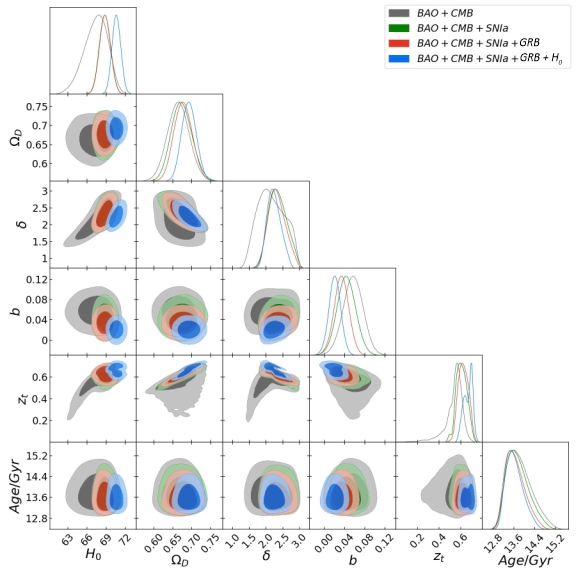}
\end{tabular}
\caption{\footnotesize The contour maps of the non-interacting THDE with consideration of the Hubble horizon as IR cutoff. In this figure $H_0$ is the Hubble parameter, $\Omega_D$ is the dark energy density, $\delta$ is the free parameter of THDE model, $z_t$ is the transition redshift and $Age/Gyr$ is the age of the universe. The best fitted values of these parameters are listed in the Table \ref{best}. The used data are BAO, CMB, Pantheon SNIa, $H_0$ and GRB. } \label{conth}
\end{figure*}

\begin{figure*}[!]
   \centering
\begin{tabular}{c}
\hspace*{-0.1in}
\includegraphics[width=0.45\textwidth]{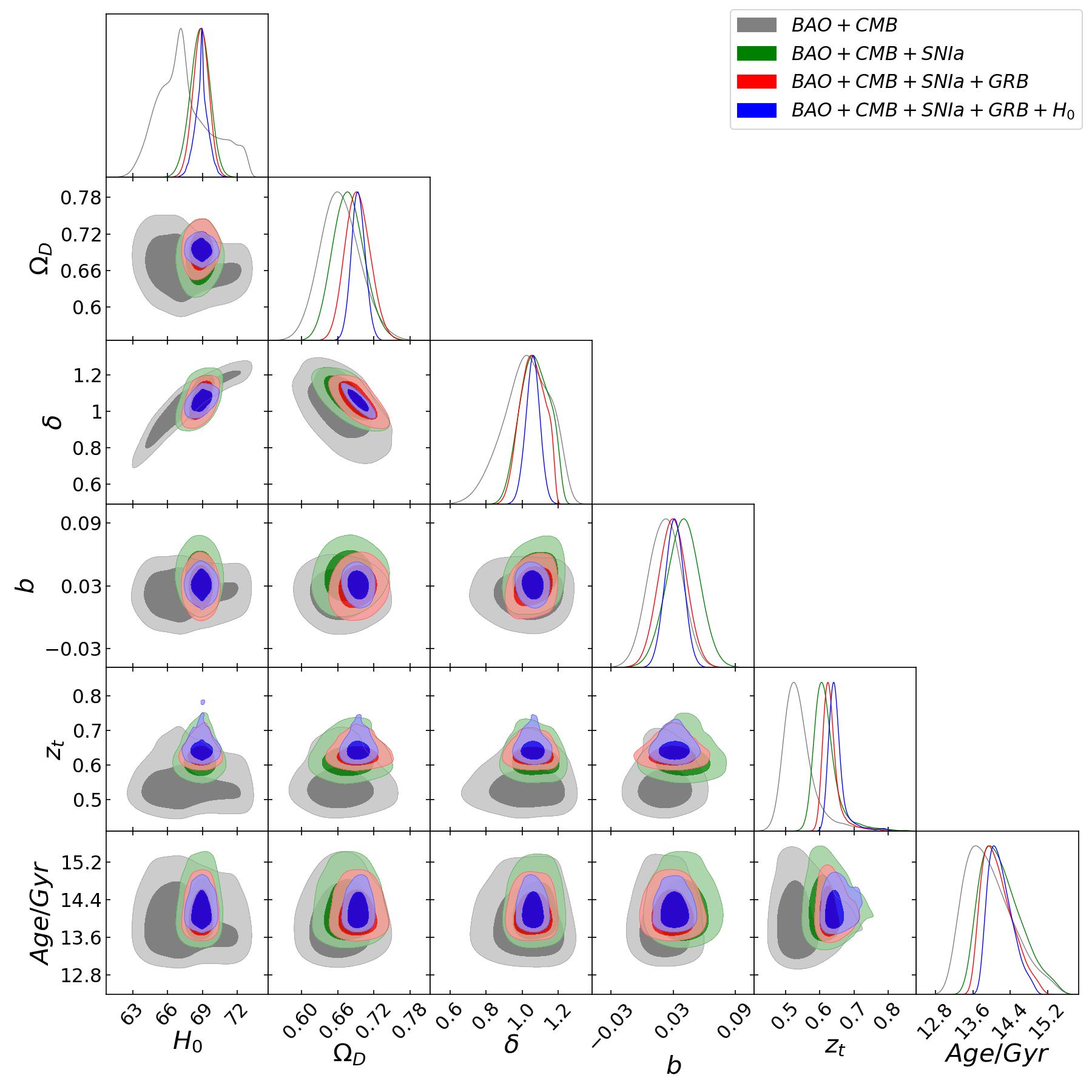}\hspace{6mm}
\includegraphics[width=0.45\textwidth]{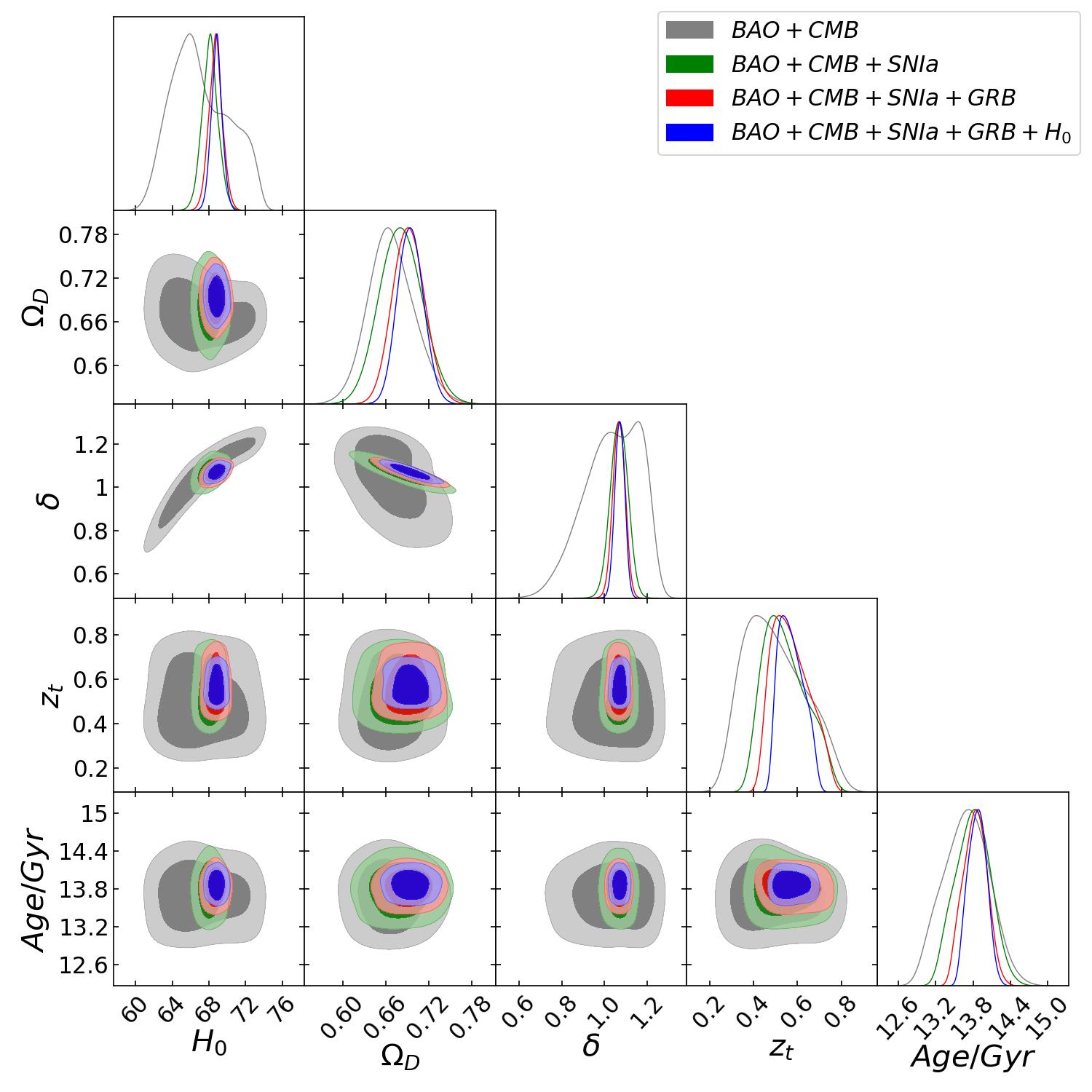}
\end{tabular}
\caption{\footnotesize The contour maps of the non-interacting THDE with consideration of the future event horizon as IR cutoff. In this figure $H_0$ is the Hubble parameter, $\Omega_D$ is the dark energy density, $\delta$ is the free parameter of THDE model, $z_t$ is the transition redshift and $Age/Gyr$ is the age of the universe. The best fitted values of these parameters are listed in the Table \ref{best}. The used data are BAO, CMB, Pantheon SNIa, $H_0$ and GRB. } \label{contfe}
\end{figure*}

The data for BAO and CMB could be found in the online source of latest version of MontePython \footnote[1]{http://baudren.github.io/montepython.html}.
Using minimized $\chi^2_{min}$, we can constrain and obtain the best-fit values of the free parameters. 
\begin{equation}\label{chi}
\chi_{min}^2=\left(\chi_{BAO}^2+\chi_{CMB}^2\right)+\chi_{Pantheon}^2+\chi_{H_0}^2+\chi_{GRB}^2
\end{equation}
The best-fit values of $\Omega_D$, $ H_0$, $ \Omega_{rc}$, c and $b$ by consideration of the $1\sigma$ confidence level are shown in the Table \ref{best} and Figs \ref{conth} and \ref{contfe}.
\begin{table*}
	\centering
	\caption{The best value of free parameters for $\Lambda$CDM, HDE, THDE and interacting THDE (ITHDE). The pantheon Supernovae data(s), Baryon acoustic oscillations(B), cosmic microwave background(C), the local Hubble constant $H_0$(H) and Gamma-Ray burst(G) have been used. The $H$ and $FE$ index stand for the Hubble horizon and the future event horizon respectively.}
		\begin{tabular}{c c c c c c c c c}
			\hline \hline
			Model  ~& $Data set$&$H_0$ & $\Omega_D$  & $ \delta $ & $c$ & $b $&$z_t$&$Age/Gyr$\\ \hline
			 
			~$\Lambda$CDM ~&    $BC$&$~66.557^{+1.750 ~}_{-1.795 ~}~$ & $0.629^{+0.056 ~}_{-0.064 ~}~$ & $-$ &$-$& $ -$ &$-$& $~13.64^{+0.40}_{-0.40}~$\\
			
			~HDE~&                       $BC$&$~68.776^{+1.751 ~}_{-1.622 ~}~$ &  $0.673^{+0.035 ~}_{-0.035 ~}~$ & $-$ &  $~0.668^{+0.080}_{-0.079}~$ &  $-$&$0.581^{+0.080}_{-0.080}$& $~13.68^{+0.46}_{-0.47}~$\\
			
			~THDE$_{H}$~            &$BC$& $~67.300^{+2.101 ~}_{-2.101 ~}~$ & $0.692^{+0.040 ~}_{-0.040 ~}~$ & $1.871^{+0.190}_{-0.441}~$ & $-$&$ - $&$0.604^{+0.110}_{-0.035}$ & $~13.97^{+0.33}_{-0.41}~$\\
			
			~ITHDE$_{H}$ ~         &$BC$&$~67.433^{+2.101 ~}_{-1.602 ~}~$ & $0.662^{+0.058 ~}_{-0.056 ~}~$ & $2.161^{+0.361}_{-0.491}~$ &$-$& $0.048^{+0.020}_{-0.020}$&$0.503^{+0.088}_{-0.037}$ & $~13.34^{+0.34}_{-0.59}~$\\

			~THDE$_{FE}$~          &$BC$& $~67.171^{+2.601 ~}_{-4.001 ~}~$ & $0.668^{+0.029 ~}_{-0.034 ~}~$ & $1.042^{+0.170}_{-0.094}~$ & $-$&$ - $&$0.502^{+0.110}_{-0.183}$ & $~13.69^{+0.37}_{-0.38}~$\\
			
			~ITHDE$_{FE}$ ~       &$BC$&$~66.872^{+1.4 ~}_{-1.4 ~}~$ & $0.664^{+0.030 ~}_{-0.035 ~}~$ & $1.031^{+0.152}_{-0.110}~$ &$-$& $0.022^{+0.016}_{-0.016}$&$0.541^{+0.023}_{-0.053}$ & $~13.95^{+0.36}_{-0.65}~$\\

			\hline \hline
			~$\Lambda$CDM ~     &$BCS$&$~68.501^{+0.849 ~}_{-0.829 ~}~$     & $0.693^{+0.020 ~}_{-0.020 ~}~$ & $-$ &$-$& $ -$ &$-$& $~13.99^{+0.31}_{-0.31}~$ \\
			
			~HDE~                        & $BCS$ &$~68.875^{+0.854 ~}_{-0.720 ~}~$   &  $0.675^{+0.023 ~}_{-0.023 ~}~$ & $-$ &  $~0.666^{+0.072}_{-0.069}~$ &  $~-~$&$0.621^{+0.034}_{-0.044}$& $~13.69^{+0.41}_{-0.40}~$ \\
			
			~THDE$_{H}$~          &$BCS$& $~69.521^{+0.802 ~}_{-0.921 ~}~$    & $0.689^{+0.039 ~}_{-0.039 ~}~$ & $2.121^{+0.150}_{-0.229}~$  & $-$ &$-$ &$0.653^{+0.044}_{-0.024}$ & $~13.96^{+0.30}_{-0.37}~$\\
			
			~ITHDE$_{H}$ ~       &$BCS$&$~68.855^{+0.871 ~}_{-0.871 ~}~$      & $0.671^{+0.055 ~}_{-0.051 ~}~$ & $2.343^{+0.252}_{-0.311}~$ &$-$& $~0.039^{+0.018}_{-0.018}~$ &$0.555^{+0.045}_{-0.054}$ & $~13.41^{+0.31}_{-0.52}~$ \\

                               ~THDE$_{FE}$~        &$BCS$& $~68.133^{+0.890 ~}_{-0.910 ~}~$      & $0.681^{+0.030 ~}_{-0.030 ~}~$ & $1.068^{+0.041}_{-0.041}~$ & $-$&$ - $&$0.545^{+0.073}_{-0.130}$ & $~13.80^{+0.28}_{-0.28}~$\\
			
			~ITHDE$_{FE}$ ~     &$BCS$&$~68.782^{+0.842 ~}_{-0.841 ~}~$       & $0.677^{+0.024 ~}_{-0.028 ~}~$ & $1.072^{+0.078}_{-0.078}~$ &$-$& $0.039^{+0.016}_{-0.016}$&$0.619^{+0.017}_{-0.040}$ & $~14.19^{+0.31}_{-0.55}~$\\
 			\hline \hline
                                ~$\Lambda$CDM ~   &$BCSG$&$~68.498^{+0.839 ~}_{-0.842 ~}~$     & $0.694^{+0.019 ~}_{-0.021 ~}~$ & $-$ &$-$& $ -$ &$-$& $~14.00^{+0.29}_{-0.29}~$  \\
			
			~HDE~                       & $BCSG$&$~68.999^{+0.551 ~}_{-0.422 ~}~$    &  $0.675^{+0.018 ~}_{-0.018 ~}~$ & $-$ &  $~0.660^{+0.067}_{-0.059}~$ &  $~-~$ &$0.633^{+0.019}_{-0.019}$& $~13.67^{+0.37}_{-0.37}~$\\
			
			~THDE$_{H}$~         &$BCSG$& $~68.752^{+0.820 ~}_{-0.820 ~}~$    & $0.691^{+0.038 ~}_{-0.038 ~}~$ & $2.054^{+0.181}_{-0.249}~$ &$ - $&$ -$&$0.677^{+0.038}_{-0.026}$ & $~13.90^{+0.26}_{-0.33}~$  \\
			
			~ITHDE$_{H}$ ~      & $BCSG$&$~68.861^{+0.865 ~}_{-0.865 ~}~$    & $0.676^{+0.047 ~}_{-0.046 ~}~$ & $2.290^{+0.220}_{-0.293}~$ &$-$& $~0.032^{+0.015}_{-0.015}~$ &$0.577^{+0.047}_{-0.047}$ & $~13.51^{+0.27}_{-0.46}~$ \\

                                ~THDE$_{FE}$~       &$BCSG$& $~68.742^{+0.750 ~}_{-0.750 ~}~$    & $0.692^{+0.022 ~}_{-0.022 ~}~$ & $1.068^{+0.028}_{-0.028}~$ & $-$&$ - $&$0.570^{+0.062}_{-0.110}$ & $~13.83^{+0.19}_{-0.19}~$\\
			
			~ITHDE$_{FE}$ ~     & $BCSG$&$~68.889^{+0.682 ~}_{-0.680 ~}~$    &$0.693^{+0.079 ~}_{-0.022 ~}~$ & $1.058^{+0.066}_{-0.066}~$ &$-$& $0.027^{+0.013}_{-0.013}$&$0.633^{+0.011}_{-0.011}$ & $~14.14^{+0.22}_{-0.42}~$\\
			\hline \hline
                                 ~$\Lambda$CDM ~  & $BCSGH$&$~69.182^{+0.788 ~}_{-0.785 ~}~$  &$0.707^{+0.019 ~}_{-0.019 ~}~$ &$-$ &$-$& $ -$ &$-$& $~13.93^{+0.25}_{-0.25}~$ \\
			
			~HDE~                      & $BCSGH$&$~69.052^{+0.501 ~}_{-0.401 ~}~$  &$0.684^{+0.011 ~}_{-0.011 ~}~$ & $-$ &  $~0.675^{+0.047}_{-0.039}~$ &  $~-~$ &$0.652^{+0.018}_{-0.017}$& $~13.75^{+0.32}_{-0.29}~$\\
			
			~THDE$_{H}$~        &$BCSGH$& $~69.664^{+0.642 ~}_{-0.733 ~}~$  &$0.690^{+0.031 ~}_{-0.031 ~}~$ & $2.211^{+0.121}_{-0.181}~$ & $-$ &$ -$&$0.670^{+0.034}_{-0.034}$ & $~13.90^{+0.23}_{-0.29}~$\\
			
			~ITHDE$_{H}$~       & $BCSGH$&$~69.532^{+0.691 ~}_{-0.691 ~}~$  &$0.695^{+0.038 ~}_{-0.038 ~}~$ & $2.244^{+0.182}_{-0.241}~$ &$-$ & $~0.022^{+0.012}_{-0.012}~$ & $0.634^{+0.051}_{-0.045}$&$~13.71^{+0.24}_{-0.41}~$  \\

                                ~THDE$_{FE}$~      &$BCSGH$& $~68.851^{+0.600 ~}_{-0.600 ~}~$   &$0.695^{+0.018 ~}_{-0.018 ~}~$ & $1.071^{+0.023}_{-0.023}~$ & $-$&$ - $&$0.572^{+0.042}_{-0.073}$ & $~13.87^{+0.15}_{-0.15}~$\\
			
			~ITHDE$_{FE}$ ~&  $BCSGH$&  $~68.930^{+0.580 ~}_{-0.581 ~}~$   &$0.694^{+0.012 ~}_{-0.012 ~}~$ &   $1.059^{+0.040}_{-0.040}~$ &  $-$&   $0.031^{+0.009}_{-0.009}$&  $0.649^{+0.010}_{-0.025}$ &   $~14.20^{+0.18}_{-0.32}~$\\
			\hline \hline
		\end{tabular}\label{best}
\end{table*}
The $\chi^2$ is known as the effective way of understanding the best values of free parameters, but it cannot be only used to determine the best model between variety of models. Hence, for this issue Akaike Information Criterion (AIC) \cite{aic} and Bayesian Information Criterion (BIC) \cite{bic} have been proposed. For further information see \cite{ab1,ab2,ab3,ab4}.
The AIC can be explained as
\begin{equation}\label{aiceq}
AIC=-2\ln\mathcal{L}_{max}+ 2k,
\end{equation}
where $-2\ln\mathcal{L}_{max}=\chi^2_{min}$ is the highest likelihood, $k$ is the number of free parameters (2 for $\Lambda$CDM, 3 for THDE and 4 for ITHDE models in addition of one further parameter $M$ for SNIa) and N is the number of data points used in the analysis. The BIC is similar to AIC with different second term
\begin{equation}\label{biceq}
BIC=-2\ln\mathcal{L}_{max}+ k\ln N.
\end{equation}
Using these definitions, it is obvious that a model giving  a small AIC and a small BIC is favored by the observations. Hence, we explain the levels of supporting the models from AIC and BIC in Tables \ref{AIC} and \ref{BIC}, respectively.\\

\begin{table}
 \centering
 \caption{ The level of support for each model from AIC.}
\begin{tabular}{c  c  }
\hline \hline
Measurment  & Explanation   \\
 \hline
  $AIC<2$& Strong support     \\
 \hline
  $2<AIC<4$ & Avarage support   \\
 \hline 
  $4<AIC<7$ &Less support   \\
  \hline
  $8<AIC$ & No support   \\
  \hline \hline
\end{tabular}\label{AIC}
\end{table}
\begin{table}
 \centering
 \caption{
The level depiction of evidence against models from BIC.}
\begin{tabular}{c  c  }
\hline \hline
Measurment  & Explanation   \\
 \hline
  $BIC<2$& No significant evidence     \\
 \hline
  $2<BIC<6$ & positive evidence   \\
 \hline 
  $6<BIC<10$ & Strong evidence    \\
  \hline
  $10<BIC$ & Very strong evidence   \\
  \hline \hline
\end{tabular}\label{BIC}
\end{table}


\section{Results}
In this section we discuss the implication of observational data for the Tsallis holographic dark energy model. Both interacting and non-interacting THDE models with assumption of the Hubble horizon and the future event horizon as IR cutoffs confronted initially with BAO and CMB and then combined with SNIa, $H_0$ and Gamma-Ray burst data. The values of cosmological parameters of the models are shown in Table \ref{best} and Figs. \ref{conth} and \ref{contfe}. We present the analysis of data in two the following parts: the cosmological parameters and the AIC and BIC model selection.
\begin{table*}
	\centering
	\caption{Summary of the AIC and BIC values calculated for interacting and non-interacting THDE model with respect to the reference $\Lambda$CDM model. $\Delta AIC=AIC_i-AIC_{\Lambda CDM}$ and $\Delta BIC=BIC_i-BIC_{\Lambda CDM}$  in which $i$ denotes the number of models $\{i=1,2,...,N\}$ with $N=5$ interacting THDE, $N=4$ for non-interacting THDE and $N=3$ for the $\Lambda$CDM models. Here we have 1169 data points. The $H$ and $FE$ index stand for the Hubble horizon and the future event horizon respectively.}
	\begin{tabular}{c  c  c c c c c}
		\hline \hline
 Data set  & $Model$ &$\chi^2_{\rm min}$ &${\rm AIC}$&$\Delta {\rm AIC}$ & ${\rm BIC}$&$\Delta {\rm BIC}$ \\\hline
$BAO+CMB$                                  &    $\Lambda$CDM &~2.037~       & ~6.037~      & ~0~ & ~6.431~       &~0~\\
 
$BAO+CMB+SNIa$                       &    $\Lambda$CDM &~1030.236~ & ~1036.236~ &~0~  &~1051.125~  &~0~\\
 
$BAO+CMB+SNIa+GRB$             &  ~  $\Lambda$CDM  ~& ~1097.615~& ~1103.615~ & ~0~ & ~1118.799~ &~0~ \\   
 $BAO+CMB+SNIa+GRB+H_0$   & ~$\Lambda$CDM~&~1097.918~ & ~1103.918~ & ~0~ & ~1118.835~ &~0~ \\
  \hline \hline
 
$BAO+CMB$  & THDE$_{H}$ &~4.833~ & ~10.833~& ~4.796~ & ~11.424~ &  ~4.99~\\
 
$BAO+CMB+SNIa$ &THDE$_{H}$ &~1032.933~ & ~1040.933~&~4.696~&~1060.785~&  ~9.66~\\
 
$BAO+CMB+SNIa+GRB$  &THDE$_{H}$ & ~1100.731~& ~1108.731~& ~5.115~ & ~1128.976~ &  ~10.177~ \\   
 $BAO+CMB+SNIa+GRB+H_0$           & ~THDE$_{H}$~        &~1104.699~      & ~1112.699~ & ~8.871~    & ~1132.947~   &  ~14.111~ \\
  \hline \hline
$BAO+CMB$                                         &ITHDE$_{H}$             &~4.429~            &~12.296~      & ~6.259~    &~13.085~        &~6.654~\\
   
$BAO+CMB+SNIa$                              &ITHDE$_{H}$              &~1031.957~      &~1041.957~ &~5.721~      &~1066.773 ~   & ~9.660~\\

$BAO+CMB+SNIa+GRB$                     &~ITHDE$_{H}$~        & ~1100.439~     & ~1110.439~ & ~5.116~     & ~1135.746 ~ &  ~16.947~\\  
 $BAO+CMB+SNIa+GRB+H_0$           & ~ITHDE$_{H}$~       &~1104.646~      & ~1114.646~ & ~10.728~   &~ 1139.957~  & ~21.122 ~\\
  \hline \hline
$BAO+CMB$                                         & THDE$_{FE}$            &~4.524~            & ~10.524~     & ~4.486~     & ~11.115~     &  ~4.684~\\
 
$BAO+CMB+SNIa$                              &THDE$_{FE}$              &~1032.438~      & ~1040.438~ &~4.201~     &~1060.291~   &  ~9.165~\\
 
$BAO+CMB+SNIa+GRB$                     &THDE$_{FE}$             & ~1100.886~     & ~1108.886~  & ~5.270~    & ~1129.131~&  ~10.332~ \\   
 $BAO+CMB+SNIa+GRB+H_0$           & ~THDE$_{FE}$ ~      &~1104.119~      & ~1112.119~  & ~8.201~  & ~1132.009~  &  ~13.193~ \\
  \hline \hline
$BAO+CMB$                                         &ITHDE$_{FE}$            &~4.420~           &~12.420~         & ~6.383~    &~13.209~     &~6.778~\\
  
$BAO+CMB+SNIa$                              &ITHDE$_{FE}$             &~1032.359~     &~1042.359~     &~6.122~      &~1067.175 ~& ~16.049~\\
  
$BAO+CMB+SNIa+GRB$                    &~ITHDE$_{FE}$ ~        & ~1100.288~ & ~1110.288~     & ~6.673~      & ~1135.595 ~&  ~16.796~\\  
 $BAO+CMB+SNIa+GRB+H_0$         & ~ITHDE$_{FE}$ ~         &~1103.420~ & ~1113.420~     & ~9.502~     &~ 1138.283~ & ~19.447 ~\\
  \hline \hline
\end{tabular}\label{aicbiclcdm}
\end{table*}

\begin{table*}
	\centering
	\caption{Summary of the AIC and BIC values calculated for interacting and non-interacting THDE model with respect to the reference HDE model. $\Delta AIC=AIC_i-AIC_{HDE}$ and $\Delta BIC=BIC_i-BIC_{HDE}$  in which $i$ denotes the number of models $\{i=1,2,...,N\}$ with $N=5$ interacting THDE, $N=4$ for non-interacting THDE and $N=4$ for the HDE models. Here we have 1169 data points. The $H$ and $FE$ index stand for the Hubble horizon and the future event horizon respectively.}
	\begin{tabular}{c  c  c c c c c}
		\hline \hline
 Data set  & $Model$ &$\chi^2_{\rm min}$ &${\rm AIC}$&$\Delta {\rm AIC}$ & ${\rm BIC}$&$\Delta {\rm BIC}$ \\\hline
$BAO+CMB$                                          & HDE &~4.315~ & ~10.315~&~0~ & ~10.906~ &~0~\\
 
$BAO+CMB+SNIa$                               &HDE                          &~1031.878~ & ~1039.878~&~0~&~1059.730~&~0~\\
 
$BAO+CMB+SNIa+H_0$                      &HDE                          & ~1101.184~& ~1109.184~&~0~ & ~1129.429~ & ~0~ \\   
 $BAO+CMB+SNIa+H_0+GRB$            & ~HDE~                   &~1106.983~ & ~1114.983~ & ~0~& ~1135.231~ &~0~ \\
  \hline \hline
 
$BAO+CMB$                                          & THDE$_{H}$          &~4.833~ & ~10.833~& ~0.518~ & ~11.424~ &  ~0.517~\\
 
$BAO+CMB+SNIa$                               &THDE$_{H}$           &~1032.933~ & ~1040.933~&~1.055~&~1060.785~&  ~1.055~\\
 
$BAO+CMB+SNIa+GRB$                      &THDE$_{H}$           & ~1100.731~& ~1108.731~& ~-0.453~ & ~1128.976~ &  ~-0.453~ \\   
 $BAO+CMB+SNIa+GRB+H_0$            & ~THDE$_{H}$~     &~1104.699~ & ~1112.699~ & ~-2.284~& ~1132.947~ &  ~-2.284~ \\
  \hline \hline
$BAO+CMB$                                         &ITHDE$_{H}$             &~4.429~            &~12.296~      & ~1.944~    &~13.085~        &~2.179~\\
   
$BAO+CMB+SNIa$                              &ITHDE$_{H}$              &~1031.957~      &~1041.957~ &~2.079~      &~1066.773 ~   & ~7.043~\\

$BAO+CMB+SNIa+GRB$                     &~ITHDE$_{H}$~        & ~1100.439~     & ~1110.439~ & ~1.255~     & ~1135.746 ~ &  ~6.317~\\  
 $BAO+CMB+SNIa+GRB+H_0$           & ~ITHDE$_{H}$~       &~1104.646~      & ~1114.646~ & ~-0.337~   &~ 1139.957~  & ~2.274 ~\\
  \hline \hline
$BAO+CMB$                                          & THDE$_{FE}$      &~4.524~ & ~10.524~& ~0.208~ & ~11.115~ &  ~0.208~\\
 
$BAO+CMB+SNIa$                               &THDE$_{FE}$        &~1032.438~ & ~1040.438~&~0.560~&~1060.291~&  ~0.560~\\
 
$BAO+CMB+SNIa+GRB$                      &THDE$_{FE}$        & ~1100.886~& ~1108.886~& ~-0.298~ & ~1129.131~ &  ~-0.701~ \\   
 $BAO+CMB+SNIa+GRB+H_0$            & ~THDE$_{FE}$~  &~1104.119~ & ~1112.119~ & ~-2.864~& ~1132.009~ &  ~-3.22~ \\
  \hline \hline
$BAO+CMB$                                          &ITHDE$_{FE}$      &~4.420~ &~12.420~ & ~2.105~ &~13.209~ &~2.309~\\
 
$BAO+CMB+SNIa$                               &ITHDE$_{FE}$       &~1032.359~ &~1042.359~ &~2.48~ &~1067.174 ~& ~7.44~\\
  
$BAO+CMB+SNIa+GRB$                      &~ITHDE$_{FE}$~ & ~1100.288~ & ~1110.288~ & ~1.104~ & ~1135.594 ~&  ~6.164 ~\\  
 $BAO+CMB+SNIa+GRB+H_0$            & ~ITHDE$_{FE}$~&~1103.420~ & ~1113.420~& ~-1.562~ &~ 1138.283~ & ~3.051 ~\\
  \hline \hline
\end{tabular}\label{aicbichde}
\end{table*}

\textbf {Cosmological Parameters}; As a key factor in modern cosmology for calculating the age and the size of the Universe and consideration of this quantity for measuring the brightness and the mass of stars, the Hubble constant $H_0$ is of utmost importance. The Hubble constant corresponds to the Hubble parameter at the observation time. According to the best fitted value of the Hubble parameter using latest observational data in this work, we found that the Hubble parameter for the interacting and non-interacting THDE models constrained from BAO and CMB is close to the obtained value of the Hubble parameter from the Planck mission ($H_0=67.66\pm0.42$)\citep{plankmission}, DES collaboration ($H_0=67.77\pm1.30$) \citep{h02} and SDSSIII BOSS ($H_0=67.60\pm0.7$)\citep{h03}. Adding SNIa, $H_0$ and GRB result in the bigger value of Hubble parameter. It is observed that the error bars of the Hubble parameter using the CMB and BAO are remarkably large and adding each data set makes the constraints to be narrower. It can be seen that the value of the dark energy for interacting THDE is smaller than the non-interacting model once we imply the BAO, CMB and SNIA for fitting parameters. After adding Gamma-Ray burst and $H_0$ the dark energy density of both interacting and non-interacting THDE are approximately identical. Totally the obtained values of dark energy density for THDE at $68\%$ confidence level is compatible with latest obtained results \citep{plankmission, cmb1,od02}. The value of the coupling constant at $1\sigma$ is obtained less than $0.1$ similar to the previous results of  interacting HDE, RDE and NHDE models\citep{b1,b2,b3,b4} and it conveys the possibility of decaying the dark energy to the dark matter.\\
Choosing the Hubble horizon as IR cutoff and with open range of prior in MCMC program, the value of $\delta$ has a tendency toward 2. This does not include the normal entropy. With this condition we can see that the value of $\delta$ once we choose the future event horizon as IR cutoff is close to 1. This is the proper result for THDE model which embraces the standard HDE and entropy.
We obtain the transition redshift using Brent's method. This method uses the inverse quadratic interpolation as a secured version of the secant algorithm and using three prior points can estimate the zero crossing\citep{zt}. The obtained values for transition redshift listed in Table\ref{best} is in range ($0.5<z_t<0.7$) and in good agreement with recent obtained result for the transition redshift \citep{zt2,zt3,zt4,zt5,zt6}(To mention few) at $1\sigma$ and $2\sigma$ interval level.

\textbf{AIC and BIC model selection}: We investigate the models according to the objective Information Criterion (IC) containing AIC (Eq.(\ref{aiceq})) and BIC (Eq.(\ref{biceq})). We present the obtained results of AIC and BIC in the Table \ref{aicbiclcdm} with consideration of the $\Lambda$CDM as the referring model. According to the values of $\Delta$AIC, $\Delta$BIC, the definition of AIC supporting area (Table. \ref{AIC}) and BIC evidence against the models (Table. \ref{BIC}), it can be seen that both ITHDE and THDE are ruled out and unsupported by observational data. The BIC imposes a strict penalty against the additional parameters more than AIC as we can see in Tables \ref{aicbiclcdm} and \ref{aicbichde}. In this case we may reject the model (THDE) as a disfavored model, but we should note that the reason of the existence of various holographic dark energy models is to alleviate the $\Lambda$CDM problems. Thus, in this work we used the holographic dark energy (HDE) model as another reference for making an accurate comparison. In this manner, we give the results of AIC and BIC in the Table\ref{aicbichde} with consideration of the HDE as the referring model. It is evident that the values of $\chi^2$, AIC and BIC for THDE are close to  the values of the $\chi^2$, AIC and BIC for HDE model and even smaller with additional GRB and $H_0$ data. But for interacting case still BIC has positive evidence against the THDE model. Thus, taking the HDE as the main model for comparison, one can see that the observational data strongly favor and support the THDE and averagely support the ITHDE model.

\begin{figure}[!]
   \centering
\begin{tabular}{c}
\hspace*{-0.1in}
\includegraphics[width=0.45\textwidth]{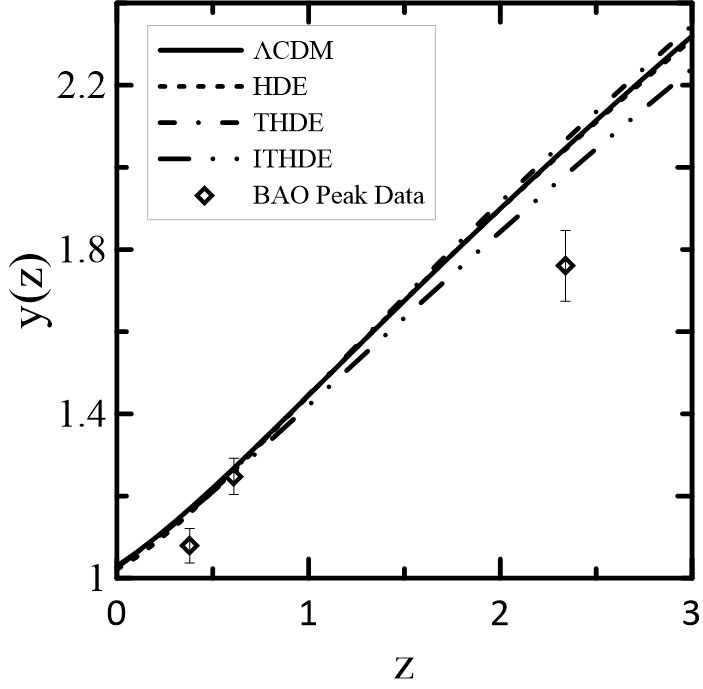}
\end{tabular}
\caption{\footnotesize  The evolution of $AP$ test versus redshift for the THDE model with consideration of the Hubble horizon as IR cutoff. The data used for plotting the trajectories of $\Lambda$CDM, HDE, THDE and ITHDE is the CMB + BAO category.} \label{apfig}
\end{figure}
\begin{figure}[!]
   \centering
\begin{tabular}{c}
\hspace*{-0.1in}
\includegraphics[width=0.45\textwidth]{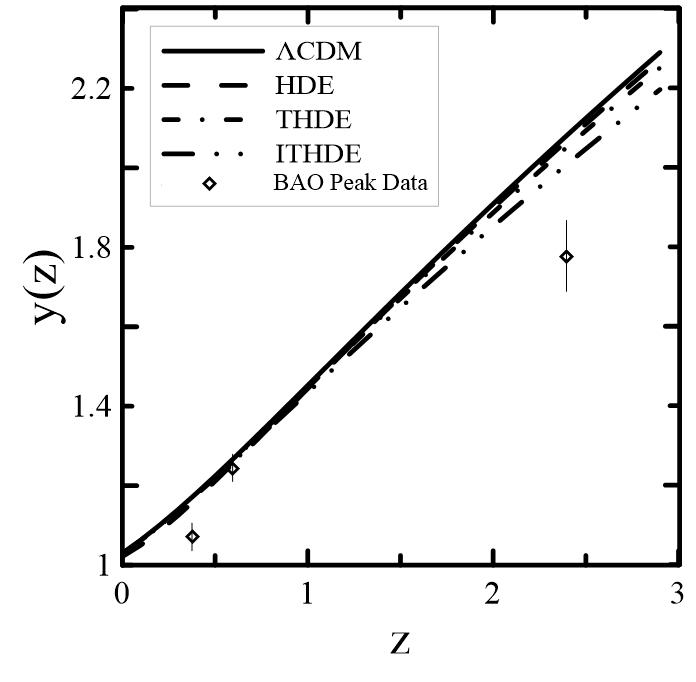}
\end{tabular}
\caption{\footnotesize  The evolution of $AP$ test versus redshift for the THDE model with consideration of the future event horizon as IR cutoff. The data used for plotting the trajectories of $\Lambda$CDM, HDE, THDE and ITHDE is the CMB + BAO category.} \label{apfigfe}
\end{figure}
\section{Alcock-Paczynski Test}
The Alcock-Paczynski (AP) test is a thoroughly geometric investigator of the cosmic expansion using observed/measured tangential and radial dimensions of objects being known as isotropic\citep{apref}. The significant advantage of this test is its independency on the galaxies' evolution. In this paper we use this method as a test for the THDE cosmological model. We also carry out the AP test according to the best fitted results using BAO, CMB. We take the $\Lambda$CDM and HDE as the reference models to compare with THDE.\\
According to the radius of objects' distribution along the line of sight
\begin{equation}\label{s11}
s_\parallel=\Delta z\frac{d}{dz}d_c(z),
\end{equation}
where $d_c$ is the comoving distance and the radius of objects' distribution perpendicular to the line of sigh
\begin{equation}\label{s1-}
s_\perp=\Delta\theta(1+z)^md_A(z),
\end{equation}
in which $\Delta z$ is the redshift span, $\Delta\theta$ is the angular size and $m=1,0$ denote the expanding and static Universe respectively, one can find the following ratio
\begin{equation}\label{y}
y\equiv \frac{\Delta z}{z\Delta\theta}\frac{s_\parallel}{s_\perp},
\end{equation}
which using the definition of the diameter angular distance if the Universe is expanding, the Eq.\ref{y} can be written as
\begin{equation}\label{yz}
y(z)=\left(1+\frac{1}{z}\right)\frac{d_A(z)H(z)}{c}.
\end{equation}

This relation is against the incorrect cosmological parameters and models. Using this relation one can check the deviation from the reference model which means the deviation from the correct measurement. In Figs.\ref{apfig} and \ref{apfigfe} we compare the THDE with HDE and $\Lambda$CDM model with considering the Hubble horizon and the future event horizon as IR cutoff respectively. This comparison has been performed using the fitted parameters of the models (See Table \ref{best}). According to the $y(0.38)=1.079\pm0.042$, $y(z=0.61)=1.248\pm0.044$ and $y(z=2.34)=1.706\pm0.083$ \citep{apmelia}the $\Lambda$CDM is not favor by the BAO data while we choose the $\Lambda$CDM as the model of comparison. The evolution of Alcock-Paczynski for $\Lambda$CDM, HDE, THDE and ITHDE in terms of redshift is plotted in Figs.\ref{apfig} and \ref{apfigfe} for THDE with the Hubble horizon and the future event horizon respectively. All models at $z=0$ have identical values of $y$. It is observed that they behave similar to $\Lambda$CDM in low redshift while in higher redshift the deviation from the reference model can be seen. The deviation of THDE and ITHDE with Hubble horizon as IR cutoff can be seen at $z=1.2$ and $z=0.7$ respectively. The deviation of THDE and ITHDE with the future event horizon as IR cutoff also can be seen at $z=1.5$ and $z=1.05$ respectively. Accordingly, it can be seen that the THDE with the future event horizon shows a small deviation respect to the $\Lambda$CDM model and also HDE model compared to the THDE with the Hubble horizon.
 \begin{figure}[!]
   \centering
\begin{tabular}{ccc}
\hspace*{-0.1in}
\includegraphics[width=0.45\textwidth]{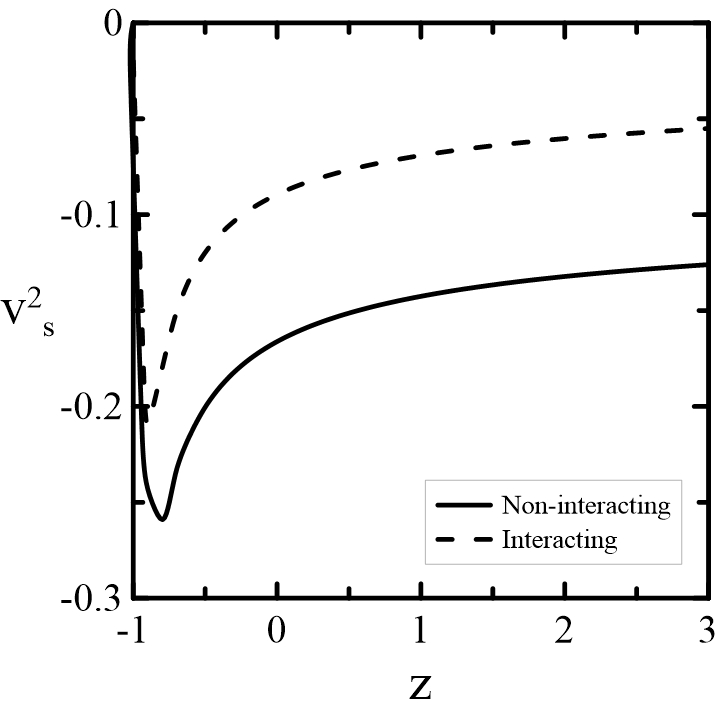}\end{tabular}
\caption{\small The evolution of $v^2_s$ versus redshift for the THDE model with consideration of the Hubble horizon as IR cutof.. Dashed line indicates the interacting and solid line indicates the non-interacting model according to the best fitted value of parameters listed in the Table \ref{best}. The negative value of trajectory shows the instability against perturbation of the background.} \label{vs}
\end{figure}
\begin{figure}[!]
   \centering
\begin{tabular}{ccc}
\hspace*{-0.1in}
\includegraphics[width=0.43\textwidth]{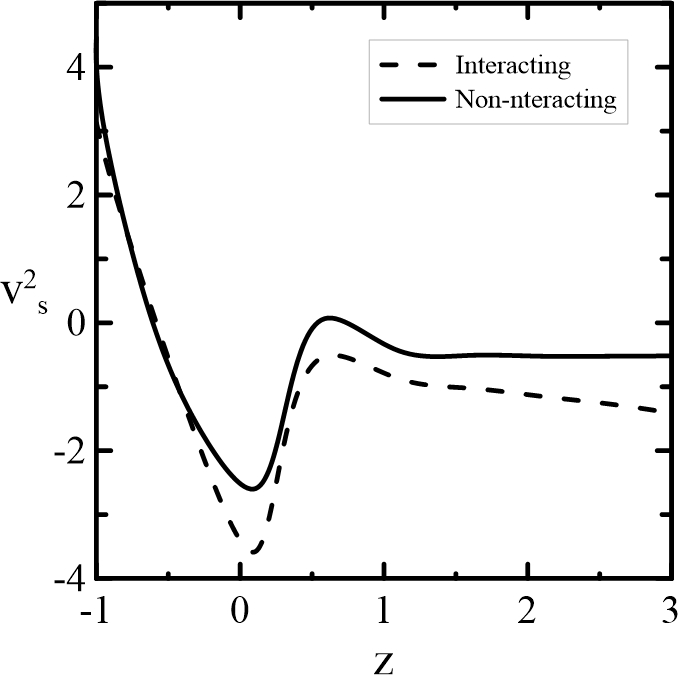}\end{tabular}
\caption{\small The evolution of $v^2_s$ versus redshift for the THDE model with consideration of the future event horizon as IR cutoff. Dashed line indicates the interacting and solid line indicates the non-interacting model according to the best fitted value of parameters listed in the Table \ref{best}. The negative value of trajectory shows the instability against perturbation of the background.} \label{vs}
\end{figure}
\section{Stability}
Surveying the stability of THDE can be performed by study the behavior of square sound speed ($v^2_s$) \cite{56}. The sign of $v_s^2$ is important to specify the stability of background evolution which $v^2_s>0$ and $v^2_s<0$ denote a stable and unstable universe against perturbation respectively. The perturbed energy density of the background in a linear perturbation structure is
\begin{equation}\label{ped}
\rho\left(x,t\right)=\rho\left(t\right)+\delta\rho\left(x,t\right),
\end{equation}
in which $\rho\left(t\right)$ is unperturbed energy density of the background. The equation of energy conservation is \cite{56}
\begin{equation}\label{ece}
\delta\ddot{\rho}=v^2_s\bigtriangledown^2\delta\rho\left(x,t\right).
\end{equation}
For positive sign of squared sound speed the Eq.\ref{ece} will be a regular wave equation which its solution can be obtained as $\delta\rho=\delta\rho_0e^{-i\omega_0t+ikx}$ indicating a propagation state for density perturbation. It is easy to see that the squared sound speed can be written as
\begin{equation}\label{vs}
v^2_s=\frac{\dot{P}}{\dot{\rho}}=\dot{\omega}_D\frac{\rho_D}{\dot{\rho}_D}+\omega_D,
\end{equation}
Taking time derivative of Eq.\ref{THDE2} and again using the Eq.\ref{THDE2} yields
\begin{equation}\label{rodro}
\frac{\rho_D}{\dot{\rho}_D}=\frac{1}{3H}\frac{((2-\delta)\Omega_D-1)}{(2-\delta)(1-\Omega_D+3b\Omega_D)},
\end{equation}
Combining the Eqs.\ref{rodro} and \ref{cons2} we have
\begin{equation}\label{wd}
\dot{\omega}_D=H\frac{(2-\delta)(1-\delta)\Omega'_D+3b(2-\delta)\Omega'_D}{(2-\delta)\Omega_D-1},
\end{equation}
Now using Eqs.(\ref{omg}), (\ref{hhd}), (\ref{vs}), (\ref{rodro}) and (\ref{wd}) we can check the stability of the THDE model. From the Fig.\ref{vs} one can see that during the cosmic evolution, both interacting and non-interacting THDE are unstable against background perturbations in early time, present and late time.\\
Taking time derivative of Eq.(\ref{THDErh}) and using Eq.(\ref{THDErh}), $\dot{R}_h=HR_h-1$ and $R_h=\frac{\rho_D}{B}^{1/(2\delta-4)}$ we have
\begin{equation}\label{rodroh}
{\left(H^2\Omega_D\right)^{\frac{2\delta-5}{2\delta-4}}\left(2(1-\delta)\Omega_DH'-H\Omega'_D\right)+3bH^4\Omega'_D}{3H^3\Omega_D^2},
\end{equation}
\begin{equation}\label{wdh}
\dot{\omega}_D=\frac{\left(3 b H^4\Omega'_D-\left(H^2 \Omega_D\right)^{\frac{5-2 \delta}{4-2 \delta}}\right)\left(2 (\delta-1) \Omega_D H'+H\Omega'_D\right)}{3 H^3 \Omega_D^2}
\end{equation}
where $H'=\dot{H}/H$ and $\Omega'_D=\dot{\Omega}_D/H$. Using Eqs. (\ref{omgRh}), (\ref{HRhp}), (\ref{vs}), (\ref{rodroh}) and (\ref{wdh}) one may investigate the evolution of stability versus redshift for the THDE model with the future event horizon as IR cutoff. From this figure we can see that the future event horizon possibles the stability of the model at the late time.
 \begin{figure}[h]
  \centering
\includegraphics[width=0.47\textwidth]{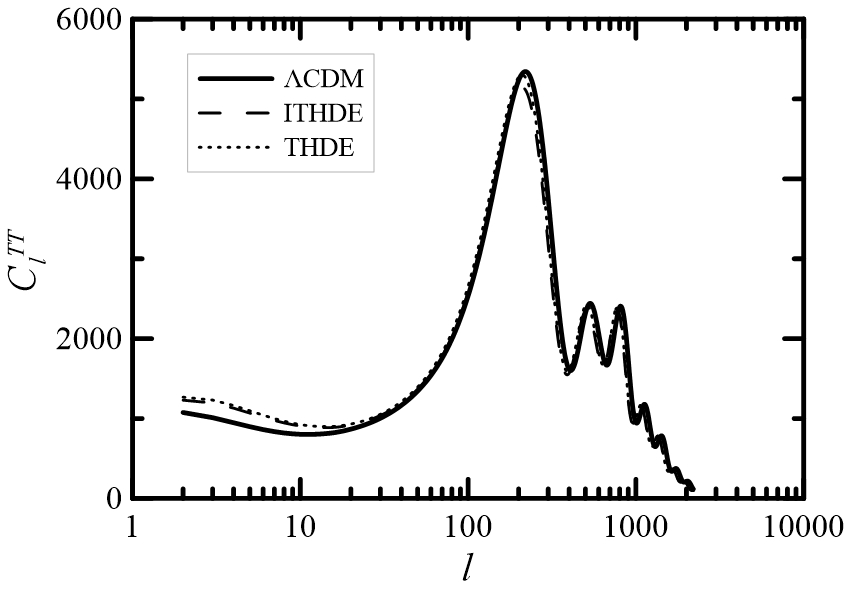}
\caption{\small The power spectrum of cosmic microwave background anisotropy of interacting and non-interacting THDE models with consideration of the Hubble horizon as IR cutoff compared to the $\Lambda$CDM. The amplitud in small $\ell$-poles for THDE is higher than the $\Lambda$CDM model. } \label{cl}
\end{figure}
 \begin{figure}[h]
   \centering
\includegraphics[width=0.47\textwidth]{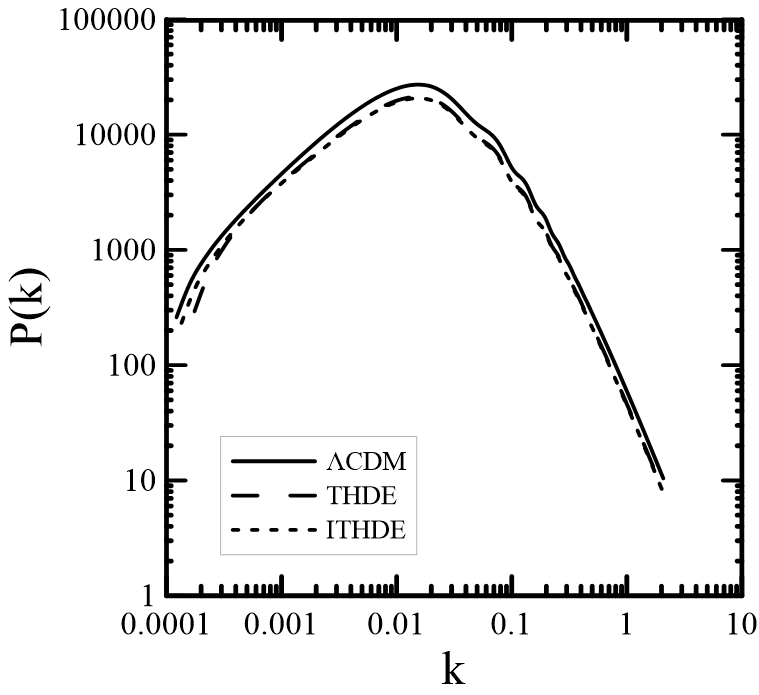}
\caption{\small The matter power spectrum of interacting and non-interacting THDE models with consideration of the Hubble horizon as IR cutoff compared to the $\Lambda$CDM according to the best fitted values listed in Table\ref{best}. Both interacting and non-interacting THDE models show a considerable suppressing of matter power spectrum at large scale or small $k$-modes.} \label{pk}
\end{figure}
 \begin{figure}[h]
  \centering
\includegraphics[width=0.47\textwidth]{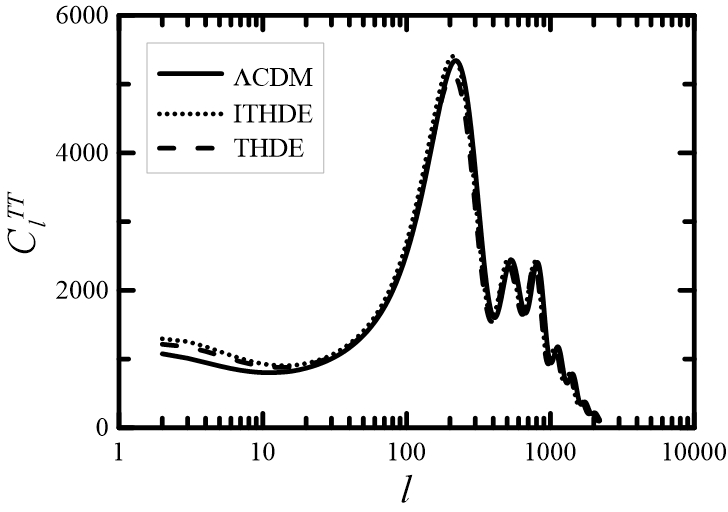}
\caption{\small The power spectrum of cosmic microwave background anisotropy of interacting and non-interacting THDE models with consideration of the future event horizon as IR cutoff compared to the $\Lambda$CDM. The amplitud in small $\ell$-poles for THDE is higher than the $\Lambda$CDM model. } \label{cl_fe}
\end{figure}
 \begin{figure}[h]
   \centering
\includegraphics[width=0.47\textwidth]{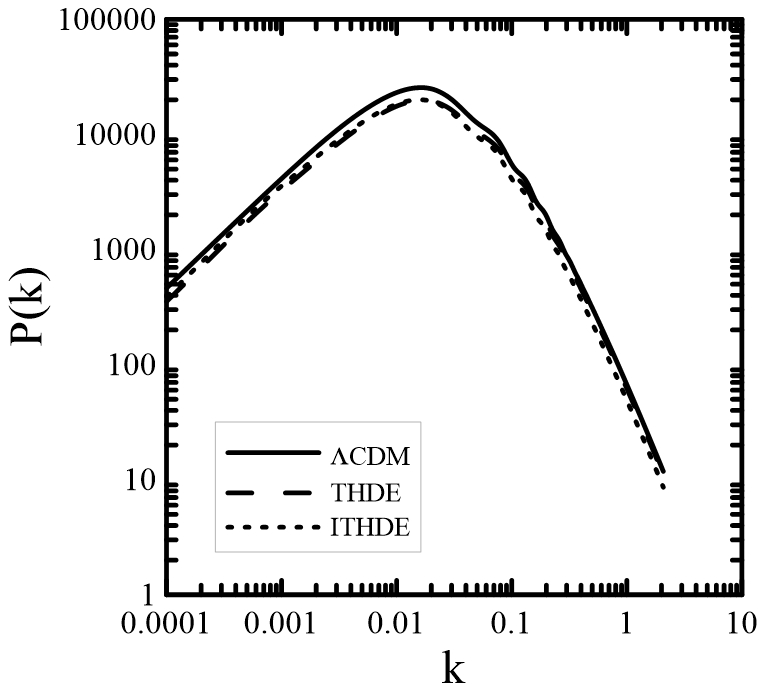}
\caption{\small The matter power spectrum of interacting and non-interacting THDE models with consideration of the future event horizon as IR cutoff compared to the $\Lambda$CDM according to the best fitted values listed in Table\ref{best}. Both interacting and non-interacting THDE models show a considerable suppressing of matter power spectrum at large scale or small $k$-modes.} \label{pk_fe}
\end{figure}
\section{CMB Power Spectrums}
In this section by the use of modified version of the Boltzmann code CAMB\footnote{https://camb.info} \cite{{camb1},{camb2}}, we compare the power spectrum of the cosmic microwave anisotropy in both interacting and non-interacting THDE models. Our results of the temperature power spectrum ($TT$) according to the fitted results in Table\ref{best} are depicted in Figs. \ref{cl} and \ref{pk} for THDE with the Hubble horizon as IR cutt of and Figs \ref{cl_fe} and \ref{pk_fe} for THDE with consideration of the future event horizon as IR cutoff. From the figures, one can see that both the interacting and non-interacting THDE models show the trends of squeezing power spectrum of the cosmic microwave anisotropy to small $\ell$ or large angle scales. This squeezing can also be seen from the power spectrum of matter distribution in the Universe. Embodying on the large scale structure of matter distributions, both models exhibit approximately $20\%$ suppressing in the peak of power spectrum which occur in small $k$ or large scale region. Another difference between the THDE and $\Lambda$CDM models lies before $\ell<50$ where the amplitude of THDE is higher than the $\Lambda$CDM. The tendency of interacting THDE model is more than THDE towards $\Lambda$CDM.
\section{Conclusion}
In this work,  using various cosmological tests we examined the Tsallis holographic dark energy model (THDE) with consideration of the Hubble horizon and the future event horizon as IR cutoffs. In this case we considered a phenomenological non-gravitational interaction between dark sectors. We used the Pantheon Supernovae type Ia, Baryon acoustic oscillation, cosmic microwave background, the local value of Hubble constant $H_0$ and the Gamma-Ray burst data as the observational data for constraining the free parameters of the models. For minimizing the $\chi^2$ we used MCMC method by employing the Cosmo Hammer (EMCEE) Python package. We observed that concerning the density of dark energy and the Hubble parameter, the THDE and ITHDE models have a good consistency with latest observational data. Both interacting and non-interacting THDE enter the accelerating universe within the $z_t=[0.5,0.7]$.  We investigated the models using the objective Information Criterion (IC) including AIC and BIC. We found that the interacting and non-interacting THDE models are not supported by observational data. This result is obtained once the $\Lambda$CDM is chosen as the reference model. According to this case that the holographic dark energy models has been proposed to alleviate the problems of $\Lambda$CDM, one can compare the THDE with another holographic models (here HDE) as the reference rather than the $\Lambda$CDM. In this case, by choosing the HDE as the referring model, both interacting and non-interacting THDE models are  favored strongly by AIC and moderately by BIC. Using Alcock-Paczynski (AP) we found that the HDE has the smallest deviation from the $\Lambda$CDM model. Accordingly both interacting and non-interacting THDE models behave similar to $\Lambda$CDM at low-$z$ but with consideration of the future event horizon as IR cutoff the deviation compared to $\Lambda$CDM and HDE can be seen at $z>1.05$ and $z>1.5$ for THDE and ITHDE respectively which is by far better that the results of the choosing the Hubble horizon. In addition Using the squared of sound speed $v^2_s$ we found that the interacting and non-interacting THDE with consideration of the Hubble horizon as IR cutoff could not satisfy the condition of stability and remain as unstable model against the background. However, the interacting and non-interacting THDE with the future event horizon show stability against the background at the late time. Finally, using modified version of CAMB package, we observed that the $20\%$ suppression of matter power spectrum from interacting and non-interacting THDE models in large scale region.\\
According to the aforementioned results, we can see that the THDE model has more acceptable behaviors in different observational tests once the future event horizon is opted as IR cutoff of the system.
It can be mentioned that for better revealing the deeper aspects of the THDE model more investigations should be done. For the future works, we would like to study the dynamical system methods for comprehension of the model's behavior in the late time using different types of interaction. In addition, we are going to to study the perturbation analysis in comparison to the Large Scale Structure (LSS) and the gravitational lenses.
\section*{ACKNOWLEDGMENTS}
The author would like to thank the referee for insightful comments which improved the quality of the paper.

\end{document}